# Gate-tunable Phase Transitions in 1T-TaS$_2$

Yijun Yu[1], Fangyuan Yang[1], Xiu Fang Lu[2], Ya Jun Yan[1,2], Y. H. Cho[3,4], Liguo Ma[1], Xiaohai Niu[1], Sejoong Kim[5], Young-Woo Son[5], Donglai Feng[1], Shiyan Li[1], Sang-Wook Cheong[3,4], Xian Hui Chen[2] and Yuanbo Zhang[1*]

[1]*State Key Laboratory of Surface Physics and Department of Physics, Fudan University, Shanghai 200438, China*

[2]*Hefei National Laboratory for Physical Science at Microscale and Department of Physics, University of Science and Technology of China, Hefei, Anhui 230026, China*

[3]*Rutgers Center for Emergent Materials and Department of Physics and Astronomy, Rutgers University, Piscataway, New Jersey 08854, USA*

[4]*Laboratory for Pohang Emergent Materials and Department of Physics, Pohang University of Science and Technology, Pohang 790-784, Korea*

[5]*Korea Institute for Advanced Study, Hoegiro 87, Dongdaemun-gu, Seoul, Korea*

*Email: zhyb@fudan.edu.cn

**The ability to tune material properties using gate electric field is at the heart of modern electronic technology[1,2]. It is also a driving force behind recent advances in two-dimensional systems, such as gate-electric-field induced superconductivity and metal-insulator transition[3–6]. Here we describe an ionic field-effect transistor (termed 'iFET'), which uses gate-controlled lithium ion intercalation to modulate the material property of layered atomic crystal 1T-TaS$_2$. The extreme charge doping induced by the tunable ion intercalation alters the energetics of various charge-ordered states in 1T-TaS$_2$, and produces a series of phase transitions in thin-flake samples with reduced dimensionality. We find that the charge-density-wave states in 1T-TaS$_2$ are three-dimensional in nature, and completely collapse in the two-dimensional limit defined by their critical thicknesses. Meanwhile the ionic gating induces multiple phase transitions from Mott-insulator to metal in 1T-TaS$_2$ thin flakes at low temperatures, with 5 orders of magnitude modulation in their resistance. Superconductivity emerges in a textured charge-density-wave state induced by ionic gating. Our method of gate-controlled intercalation of 2D atomic crystals in the bulk limit opens up new possibilities in searching for novel states of matter in the extreme charge-carrier-concentration limit.**

The competition, coexistence and cooperation of various collectively ordered electronic states in two-dimensional (2D) systems has been breeding ground for novel states of matter[7]. Controllable manipulation of the various phases through gate-tunable charge doping may lead to new device paradigm for future material science and technology. The transition metal dichalcogenide (TMD) 1T-TaS$_2$ is a layered 2D atomic crystal known to develop a peculiar Mott phase at low temperatures[8–10]. With a rich set of charge ordered states delicately balanced on a similar energy scale[8,11–13], it provides a unique opportunity for such manipulation[14–17]. The large carrier concentration already present in metallic 1T-TaS$_2$, however, poses a great challenge. Tipping the balance of various phases in 1T-TaS$_2$ requires a charge-doping level that is comparable to its initial carrier concentration, thus calls for extreme doping capability well beyond conventional gate-electric-field induced doping level in semiconductor devices, or even electron-double-layer (EDL) surface gating using ionic liquid[18–20].

We develop a new doping method that utilizes a gate electric field to drive lithium ions in and out of the layered 1T-TaS$_2$, and introduces high doping levels in each atomic layer. This is realized in a device, referred to as ionic field-effect transistor, that controls the electronic properties of a layered material (1T-TaS$_2$ in our case) by gate-controlled intercalation. In this report, we explore previously inaccessible parameter space in 1T-TaS$_2$ based iFETs following two pathways: i) reducing the dimensionality of 1T-TaS$_2$ by thinning it down to a few atomic layers, and ii) doping it by gate-controlled intercalation. Both pathways modify the energetics of the charge-density wave (CDW) phases, and dramatically modulate the electronic properties of 1T-TaS$_2$ by inducing phase transitions. In particular, superconductivity emerges inside a textured CDW phase as a result of gate-doping. Our complete phase diagram for the first time investigates the importance of dimensionality and gate-controlled ionic doping in layered atomic crystals, and provides fresh insights into the relation between superconducting phase and various other charge-ordered phases.

Bulk 1T-TaS$_2$ has a layered structure, in which each atomic layer is composed of a plane of tantalum (Ta) atoms sandwiched by sulphur (S) atoms in an octahedral arrangement[11,13]. It undergoes a series of CDW transitions as the temperature is lowered: a metallic incommensurate CDW (ICCDW) phase below 550 K, a textured nearly commensurate CDW (NCCDW) phase below 350 K, and a commensurate CDW (CCDW) phase (accompanied by a Mott phase) below 180 K (Fig. 3a, red curve). A detailed description of those phases can be found in refs.[8,11–14,21]. We then fabricated 1T-TaS$_2$ thin-flake devices using mechanical exfoliation of the bulk crystal, followed by electrode deposition (Fig. 1a, see Supplementary Information for details).

The importance of dimensionality on the CDW phases in 1T-TaS$_2$ was probed by studying pristine thin flakes with varying thicknesses down to ~ 2 nm. We discover that both CCDW/NCCDW and NCCDW/ICCDW phase transitions (manifested as sudden jumps in the temperature-dependent resistance shown in Fig. 3a) are strongly modulated by sample thickness – an effect that eluded previous attempts[22]. As the thickness was reduced, both transitions were shifted to lowered temperatures during cool-down (warm-up transitions are discussed in Supplementary Information, section VI), then suddenly vanished at critical thicknesses of ~ 10 nm and ~ 3 nm, respectively. Meanwhile an insulating behavior prevailed in all samples below 3 nm.

The observed critical thicknesses can be understood by invoking the three-dimensional nature of the CDW ordering in 1T-TaS$_2$. Early studies [12,23,24] have suggested a 13-layer (~ 7.8 nm) and 3-layer (~ 1.8 nm) periodicity in the out-of-plane direction in the CCDW and NCCDW phase, respectively. Such periodicities arise as the structure attempts to stabilize the in-plane charge orders while minimizing the sum of interlayer Coulomb and tunneling energy[24]. Our observed critical thicknesses agree reasonably well with the proposed periodicities. Along with previous results[12,23,24], our findings defy the conventional notion of 2D CDW, and point to the vital role of interlayer interaction in sustaining the long-range charge ordering.

We now turn to study 1T-TaS$_2$ thin flakes under gate modulation in an iFET configuration. A schematic of the device is shown in Fig. 1b. Gel-like solid electrolyte (LiClO$_4$ dissolved in polyethylene oxide (PEO) matrix, see Methods) was used as the gate medium covering both the thin flake and the metal side-gate. Using an 8-nm-thick flake as an example, the effect of gate bias $V_g$ is shown in Fig. 1d (sample A, black curve). We clearly observed a resistance drop to half of its original value at $V_g \sim 2$ V during gate sweep, which is identified as a doping-induced NCCDW to ICCDW phase transition (Supplementary Information, section III). Such a phase transition was not detected in samples subjected to EDL surface gating under our experimental condition (Supplementary Information, section IV), and we attribute the transition to gate-controlled intercalation which induces high electron doping in individual atomic layers of the thin flake (even though structural modification introduced by lithium ion intercalation may also play a role). This is much like the charging process in a lithium ion battery where lithium ions intercalate into the anode upon application of a voltage[25]. Indeed, 1T-TaS$_2$ crystal has been previously studied for its potential application in batteries[26], and similar gating effects were also observed in NbSe$_2$ thin films[27] and recently in graphite[28].

Further investigations support the proposed mechanism of gate-controlled intercalation in our iFET. First, we show that the iFET operates through diffusion – the gate-tunability of the sample remains unchanged even if the sample is only partially in contact with the solid electrolyte. This was demonstrated in Fig. 1d where the active area of sample B was protected by a layer of 300-nm-thick Poly(methyl methacrylate) (PMMA) with only a corner exposed to the solid electrolyte. Upon gating, sample B (red curve, Fig. 1d) behaved nonetheless the same as sample A (black curve, Fig. 1d) which was entirely exposed. Meanwhile a fully protected control sample (sample C, blue curve in Fig. 1d) showed no sign of gating effect. Our finding rules out any alternative mechanisms that require close contact between the sample surface and the electrolyte, and points to diffusion-based intercalation as the most likely scenario.

Second, we show that the 1T-TaS$_2$ iFET is capable of doping the sample far beyond the critical thickness associated with the electrostatic screening length. A gate voltage applied on the iFET induces NCCDW to ICCDW phase transition in samples with varying thicknesses from 3.5 nm to 23 nm, and some typical data sets were shown in Fig. 2a ($V_g$ is swept at 1 mV/s. Curves are shifted horizontally for clarity). Here a well-developed ICCDW phase, with a hallmark resistance factor-of-two lower than that in the initial NCCDW phase, indicates a uniform doping in the gate-induced ICCDW phase. Such a uniform ICCDW phase was clearly observed in samples with thicknesses less than 13 nm but not in thicker samples (Fig. 2a), due to limited gate-modulation depth. We were able to extract the gate-modulation depth of our iFET by simplifying the gated sample as two homogeneous slabs (Fig. 2b inset): the uniformly doped upper slab with resistance $R_0$ (low-resistance ICCDW phase), and the un-doped lower slab with resistance $2R_0$ (high-resistance NCCDW phase). The resulting lowest resistance $R$ normalized to the high resistance $2R_0$, i.e. $r = R/2R_0$, is therefore:

$$r = \begin{cases} 0.5 & \text{if } t < t_c \\ 1 - \dfrac{t_c}{2t} & \text{if } t > t_c \end{cases} \quad (1)$$

where $t$ is the sample thickness and $t_c$ the maximum gate-modulation depth. By fitting our data with above model, we obtained $t_c \approx 14$ nm (Fig. 2b). We note that even larger critical thicknesses is possible with slower gate sweep, as the lithium ions have more time to diffuse into the sample. Such a gate-modulation depth is much larger than the gating range of EDL, which is limited by the electrostatic screening length (< 1 nm), and lends further support to gate-controlled intercalation as the origin of the gate modulation. Finally we point out that the gate modulation in our iFET repeatable and reversible (Fig. 2c), suggesting that no chemical modification of the sample occurred during the intercalation cycles.

The gate-controlled intercalation dramatically modulates the various phase transitions in 1T-TaS$_2$ that were probed by measuring the temperature-dependent resistance in samples with varying thicknesses (Fig. 4). Here the temperature-dependent resistance at varying fixed $V_g$ for three sample thicknesses are presented – 14 nm in Fig. 4a, 8 nm in Fig. 4b, and 3.5 nm in Fig. 4c. These three thicknesses were chosen because they represent three distinctive regions in the phase diagram shown in Fig. 2b:

the 14-nm-thick sample retains all the CDW phases found in the bulk, so is in the bulk limit; the 8-nm-thick sample is in the quasi-2D limit, where all but the CCDW phases remain; finally in the 2D limit, the 3.5-nm-thick sample is an insulator without obvious phase transitions.

The temperature-dependent resistance of the 14-nm-thick sample is shown in Fig. 4a. For clarity the gate-induced NCCDW to ICCDW phase transition at room-temperature is used as a map (shown in the inset), where the $V_g$ is color-coded to match the values used in each temperature sweep. Upon gating, the resistance was strongly suppressed in the entire CCDW phase starting from $V_g = 2.6\,\text{V}$, even though all phase transition temperatures remained unaffected (Fig. 4a, orange curve). We attribute the observed resistance suppression to gate-induced melting of the Mott state inside of the CCDW phase. The fact that the CCDW phase survives in the absence of a Mott phase indicates that the Mott-localization is not the driving force but rather a consequence of the CCDW transition, in contrast to recent studies[29–32]. An increasing $V_g$ depressed the NCCDW/CCDW transition temperature, and eventually eliminated the transition altogether. The sample then became metallic with room-temperature resistivity per layer on the order of quantum resistance $R_Q$ defined as $h/4e^2$ (expected sample resistance corresponding to $R_Q$ is shown as dashed line in Fig. 3A as a reference. $h$ is the Plank constant and $e$ the elementary charge, respectively). A superconducting phase emerges inside of the NCCDW phase, and is characterized by a well-defined zero-resistance state below $T_c = 2\,\text{K}$ and an upper critical field of $H_{c2} \approx 0.7\,\text{T}$ (Fig. 4e and Supplementary Information, section VII). Further increase of the gate doping eliminated the ICCDW/NCCDW transition, while the superconductivity persisted to high doping levels with no appreciable change in $T_c$, before disappearing into a metallic state at $V_g \geq 3.3$ V. Fig. 5a summarizes our observations in a doping-temperature phase diagram. We note that the sharp transitions between the various phases are strong indication that the samples are homogenous, even though the exact atomic structures of the phases seen in gated flakes warrant further investigation.

The dramatic effect of gate doping is most visible at low temperatures, where the normal state resistance spans 5 orders of magnitude under gating (Fig. 4d). Three regions corresponding to CCDW, NCCDW and ICCDW phases are readily identified, and we determined the charge carrier density in the three phases from Hall measurements at $T = 10$ K (Fig. 4e, lower panel). As soon as the Mott phase was suppressed at $V_g = 2.6$ V, the CCDW state gained a carrier density that corresponds to ~ 1 $e$ per David-star (blue dashed line in Fig. 4d), as expected from a melted Mott state. Further doping significantly increased the carrier density, and eventually reaches ~ 1 $e$ per Ta atom. Such an extremely high carrier concentration is almost two orders of magnitude higher than the maximum doping expected from EDL gating [33] (red dashed line, Fig. 4d). This comparison further corroborates our proposed mechanism of gate-controlled intercalation, even if we take into account the fact that itinerate electrons freed from the melted CDW state also contribute to the carrier density[34].

The gate-controlled intercalation has a marked effect also on the 8-nm-thick sample. Here the CCDW phase (and the associated Mott phase) was absent due to the reduced dimensionality, but the remaining phases evolved in a pattern similar to that in the 14-nm-thick sample (Fig. 4b, ICCDW to NCCDW transition not shown). Meanwhile the 3.5-nm-thick sample remained insulating at low temperatures ($T < 100$ K) likely due to disorders, and gating only partially suppressed the insulating phase (Fig. 4c). These observations are summarized in the phase diagrams in Fig. 5b and 5c. Here we note that the absolute $V_g$ value does not directly quantify the extent of gate doping due to large hysteresis observed during gate sweeps (also see Supplementary Information, section IV). However, the gating behavior is robust in all our iFETs tested (41 devices in total), even though $V_g$ at the phase transitions varies from sample to sample. We instead use the carrier concentration obtained from Hall measurements as a quantitative measure of the doping level (Fig. 4d).

The new phase diagrams of 1T-TaS$_2$ provide fresh insights into the intricate relation between various charge-ordered phases and superconductivity (Fig. 5). An important clue lies in the doping-temperature phase diagram in the quasi-2D limit (Fig. 5b). Here the CCDW phase collapses due to reduced dimensionality, but all other phases, including the superconducting phase, remain intact. This observation leads us to conclude that the CCDW phase (therefore the Mott phase inside of it) is not directly responsible for the superconductivity. The question then arises as to what causes the superconductivity in 1T-TaS$_2$. To this end, we point out that the superconductivity appears at the intersection of NCCDW and ICCDW, where mesoscopic/microscopic phase separation is likely to occur due to the first-order nature of the transition[35,36]. Because the system is not superconducting deep in either phases, we speculate that the microscopically separated ICCDW/NCCDW phase at the intersection may cooperate to provide the right amount of electron-phonon interaction for Cooper pairing.

To summarize, we developed a new device referred to as iFET, and used it to controllably intercalate lithium ions into the atomic layers of 1T-$TaS_2$ by an externally applied gate voltage. Tremendous charge-doping is achieved, and we are able to modulate the electronic properties of 1T-$TaS_2$ through gate-induced phase transitions. The findings shed new light onto the intricate interplay of various phases in 1T-$TaS_2$ thin flakes. Our *in-situ* gate-controlled intercalation, as a general method for continuous, reversible tuning of material properties of 2D atomic crystals, opens up new opportunities in searching for novel materials and devices in the extreme doping limit.

**Methods**

Device fabrication

High quality 1T-TaS$_2$ single crystals were grown by standard chemical vapor transport method. 1T-TaS$_2$ thin flakes are obtained by mechanically exfoliating bulk single-crystals onto Si wafer covered with 285-nm-thick thermally grown SiO$_2$ layer (Quartz substrate was also used in some of our devices and no difference is observed in terms of device characteristics). Optical microscopy and atomic force microscopy (AFM, Park Systems XE-120) were used to characterize the quality and thickness of the thin flakes. Once thin flakes were identified on the substrate, we deposited metal electrodes (Au, typical thickness ~ 40 nm) using either standard electron-beam lithography process or direct evaporation through stencil mask. Sample degradation was observed in thin flakes stored in air (see Supplementary Information for details). Air exposure is minimized during fabrication process to mitigate the degradation of the sample.

LiClO$_4$ (Sigma Aldrich) dissolved in poly(ethylene oxide) (PEO, M$_w$=100,000, Sigma Aldrich) matrix is used as the solid electrolyte[33,37]. LiClO$_4$ and PEO powders (0.3 g and 1 g, respectively) are mixed with 15 ml anhydrous methanol (Alfa Aesar). The mixture is then stirred overnight with the temperature kept at 50 ℃. We note that the ratio between LiClO$_4$ and PEO is crucial for the performance of the electrolyte. After application of the electrolyte, all our samples were annealed at 370 K in vacuum (~ 30 Pa) for half an hour to eliminate residual methanol and moisture before the gate sweep.

Measurements

Transport measurements were mainly performed in Oxford Instruments Integra$^{\text{TM}}$ AC cryostat. Sample resistance was measured using lock-in amplifier (Stanford Research 830), and the gate voltage is supplied by Keithley 2400 source meter. We swept $V_g$ at fixed temperatures between 310 K and 370 K in vacuum. Before each cool-down, the temperature and gate voltage were held constant for half an hour to ensure samples are homogenously intercalated.

**Acknowledgements**

We thank D.-H. Lee for his critical reading of our manuscript, Z.-X. Shen, F. Wang,

L. Zhou, J. Zhao, Y. Wang, W. Wu, P. Darancet for helpful discussions, and X. Hong, L. He for assistance with low-temperature measurements. Part of the sample fabrication was conducted at Fudan Nano-fabrication Laboratory. Y.Y., F.Y., L.M. and Y.Z. acknowledge financial support from the National Basic Research Program of China (973 Program) under grant no. 2011CB921802 and 2013CB921902, and from the NSF of China under grant no. 11034001. X.F.L., Y.J.Y. and X.H.C. are supported by the "Strategic Priority Research Program (B)" of the Chinese Academy of Sciences (Grant No. XDB04040100) and the NSF of China (Grants No. 11190021). Y.-W.S. is supported by the NRF of Korea grant funded by MEST (QMMRC, No. R11-2008-053-01002-0). The work at Rutgers is supported by the NSF under Grant No. NSF-DMREF-1233349, and that at Postech is supported by the Max Planck POSTECH/KOREA Research Initiative Program (Grant No. 2011-0031558) through NRF of Korea funded by MEST.

**Figure captions**

**Figure 1. Gate-controlled intercalation in 1T-TaS$_2$ based iFET. a,** 3D rendering of AFM topography of a 1T-TaS$_2$ thin-flake device. The cross-section of the thin flake is shown in the lower panel. **b,** Schematic device structure of 1T-TaS$_2$ iFET and measurement setup. **c,** An illustration of lithium intercalated into layered 1T-TaS$_2$. **d,** Resistance of an 8-nm-thick 1T-TaS$_2$ crystal during an up-sweep of $V_g$ at 325 K (sample A, black curve). The drop in sample resistance by one half is due to the NCCDW/ICCDW phase transition described in the text. Sample A is fully exposed to solid electrolyte during measurement; a corner of sample B is exposed to solid electrolyte while the active area (the region between the electrodes) is protected by a layer of spin-coated PMMA; sample C is fully protected by PMMA. The normalized sample resistance of sample A, B and C are shown as black, red and blue curves, respectively. Scale bar: 5 μm.

**Figure 2. Gate-modulation depth and repeatability of 1T-TaS$_2$ iFET. a,** Resistance (normalized to its value in the NCCDW state) shown as a function of $V_g$ for a few typical samples with varying thicknesses. Gate-induced NCCDW to ICCDW phase transition is manifested as the sudden resistance drop by one half, and the slight upturn at high doping is due to the onset of a resistive state (not shown). Curves are shifted horizontally for clarity. **b,** The ratio $r$ between lowest resistance $R$ during gate sweep and the initial resistance $R_0$ in samples as a function of sample thickness. Red

line is a fit according to the model described in the text. Inset: inhomogeneous doing in the out-of-plane direction assumed in the model, where $t_c$ is the modulation depth of the gate-controlled intercalation. $t_c = 14$ nm is obtained from the fitting as indicated by the arrow in the main figure. **c,** Resistance of a sample during 10 gate sweep cycles. No appreciable change is observed during repeated gate-sweeps.

**Figure 3. CDW phases in pristine 1T-TaS₂ thin flakes with varying thicknesses. a,** Resistance as a function of temperature for pristine 1T-TaS₂ thin-flake samples with varying thicknesses. The red curve shows the typical behavior of a bulk crystal, and the arrows indicate the direction of temperature ramping. The first-order ICCDW/NCCDW and NCCDW/CCDW phase transitions are identified as jumps in sample resistance. **b,** Thickness-temperature phase diagram obtained from measurements in **a**. For simplicity, only transition temperatures recorded during cool-down are shown.

**Figure 4. Electronic transport in 1T-TaS₂ thin flakes under gate-controlled intercalation. a**, **b,** and **c,** Temperature-dependent resistance at fixed gate voltages for three sample thicknesses: 14 nm, 8 nm and 3.5 nm, respectively. The inset in each panel displays the gate-induced NCCDW/ICCDW phase transition at 310 K, which serves as a color-coded reference of $V_g$. **d,** The resistance and carrier density measured as a function of $V_g$ at 10 K in the same 14-nm-thick sample shown in **a**. The carrier density is obtained from Hall measurement. The red dashed line is the estimated doping level expected from electrostatic doping by EDL at sample surface [33]. Blue dashed line marks

the doping level corresponding to 1 $e$ per David-star. **e,** Zero-resistance superconducting state of a typical gated 1T-TaS$_2$ flake subjected to a varying perpendicular magnetic field.

**Figure 5. Doping-temperature phase diagrams of 1T-TaS$_2$ flakes in three thickness regimes. a**, **b,** and **c,** Doping-temperature phase diagrams of 1T-TaS$_2$ thin-flake samples in the bulk, quasi-2D, and 2D limit, respectively. They are obtained from the data shown in Fig. 3a to 3c. Only transition temperatures recorded during cool-down are shown for simplicity. The metal-insulator transition temperature in **c** is defined as the temperature where the (initially metallic) sample starts to show insulating behavior as the temperature is lowered.

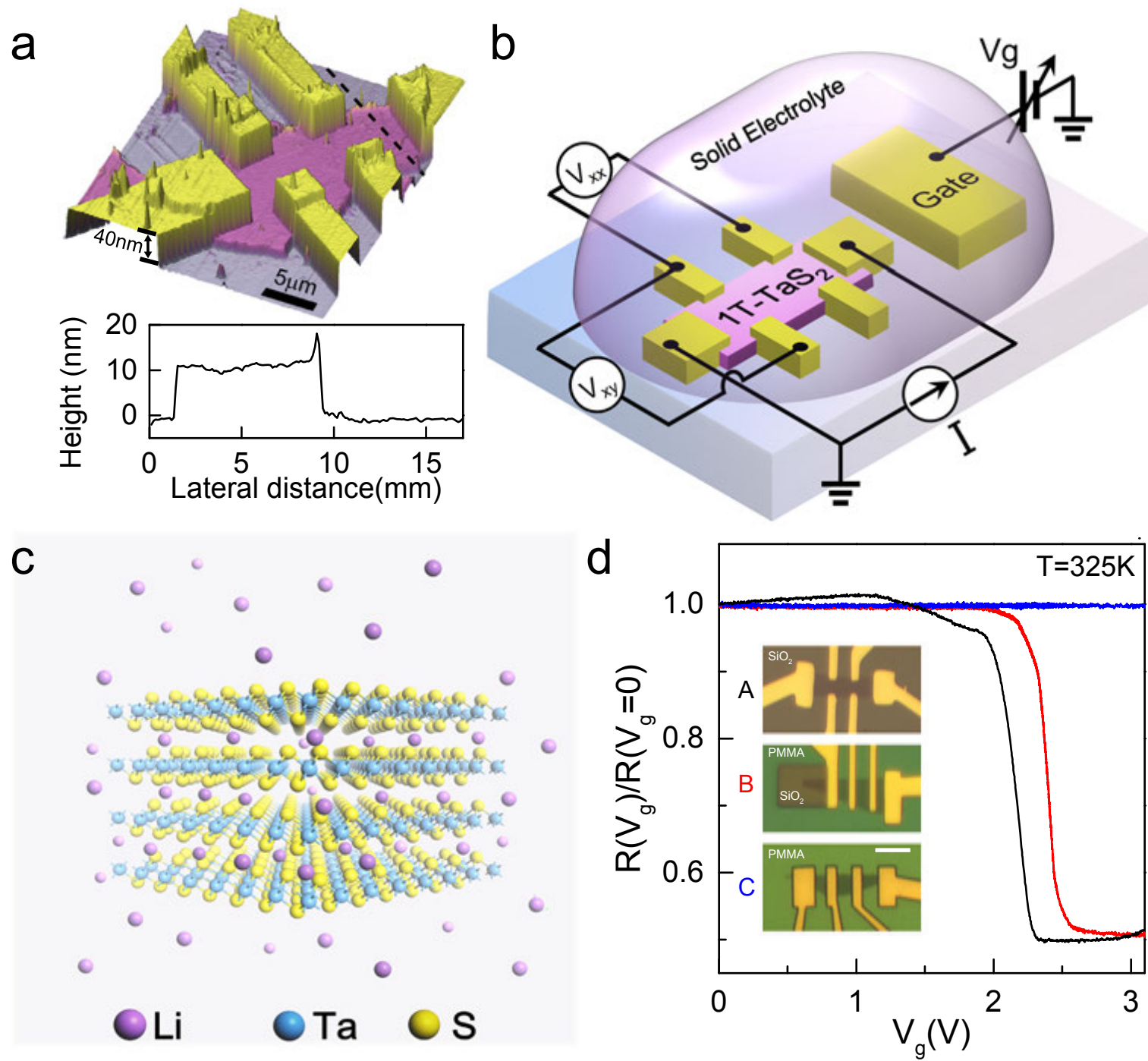

Fig. 1, Y. Yu *et al*

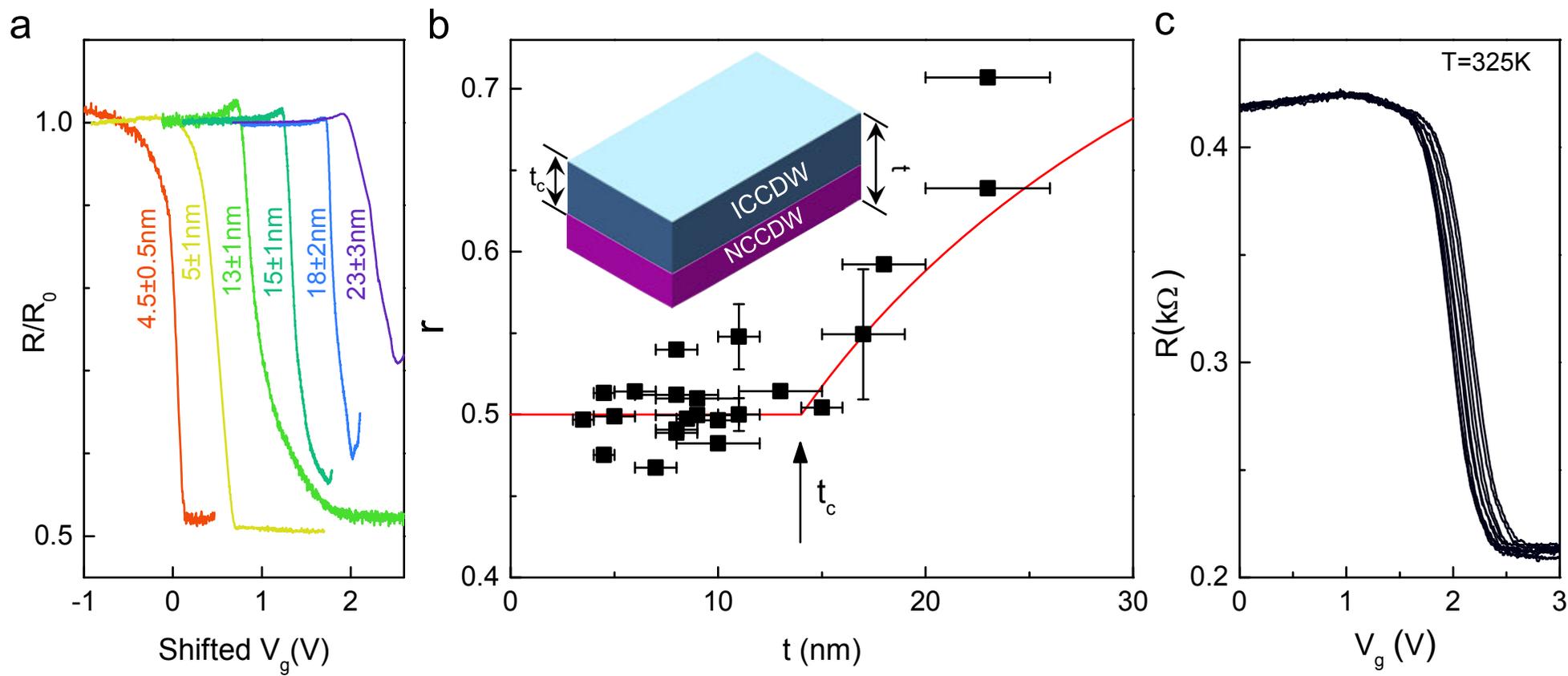

Fig. 2, Y. Yu *et al*

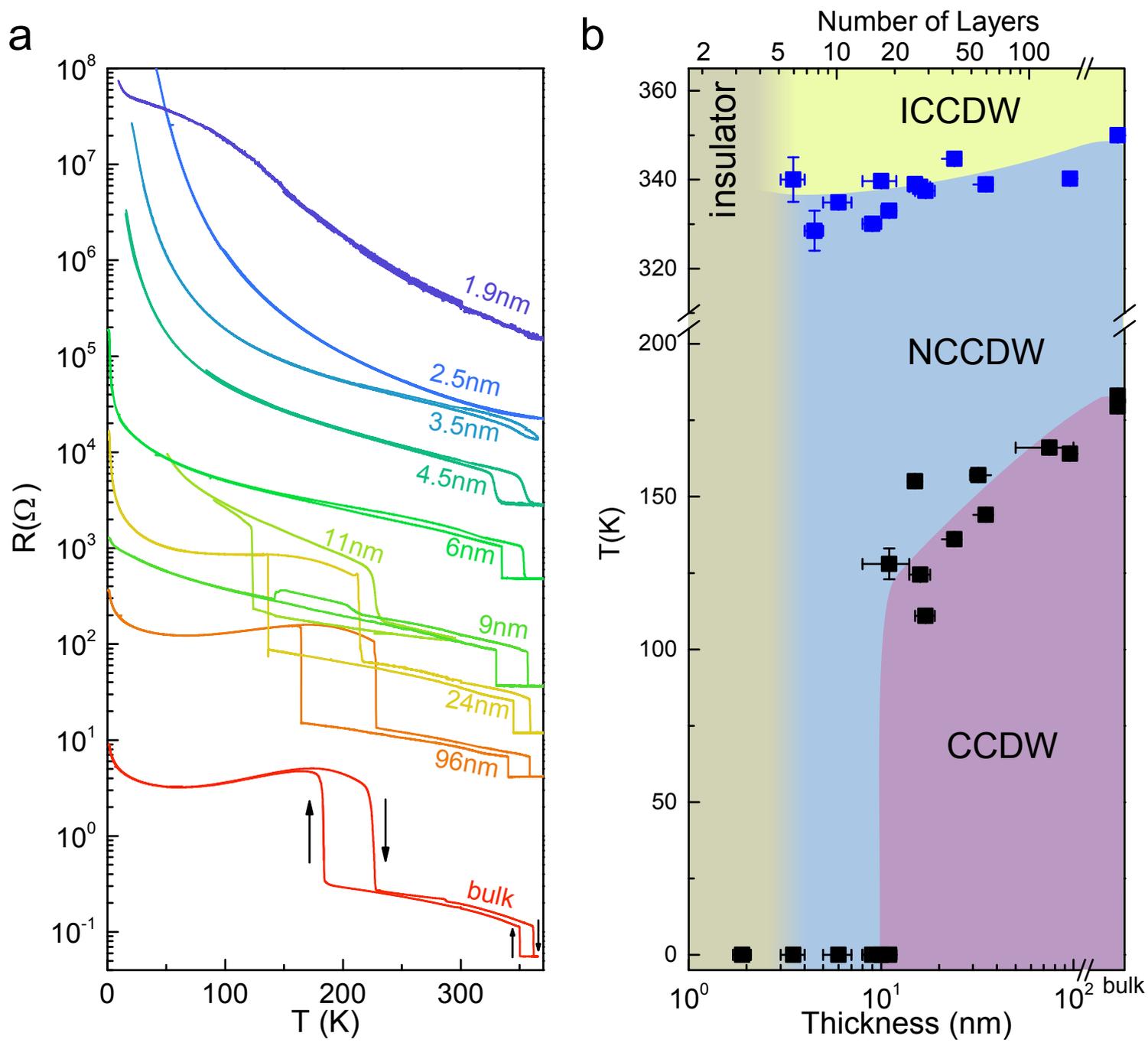

Fig. 3, Y. Yu *et al*

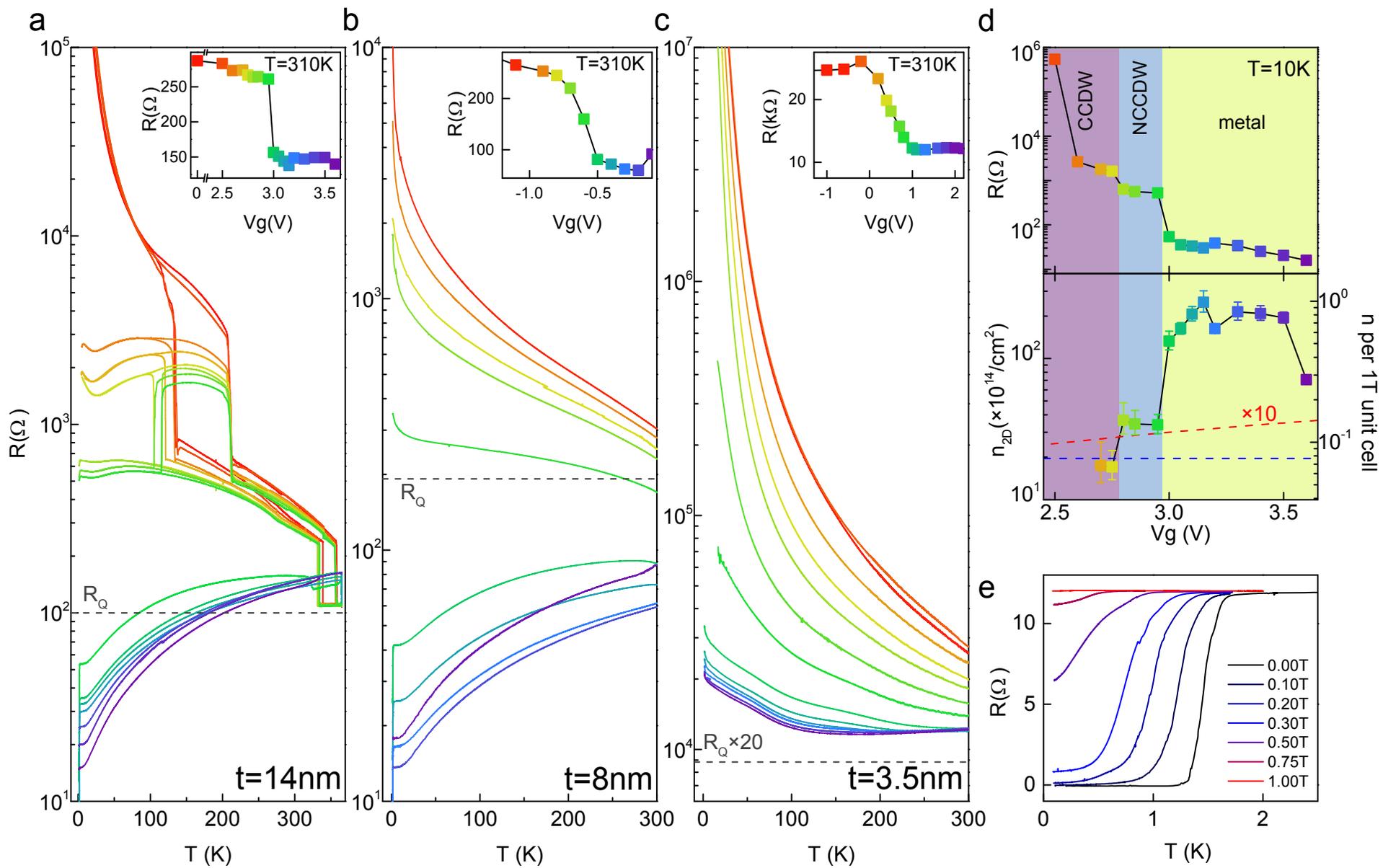

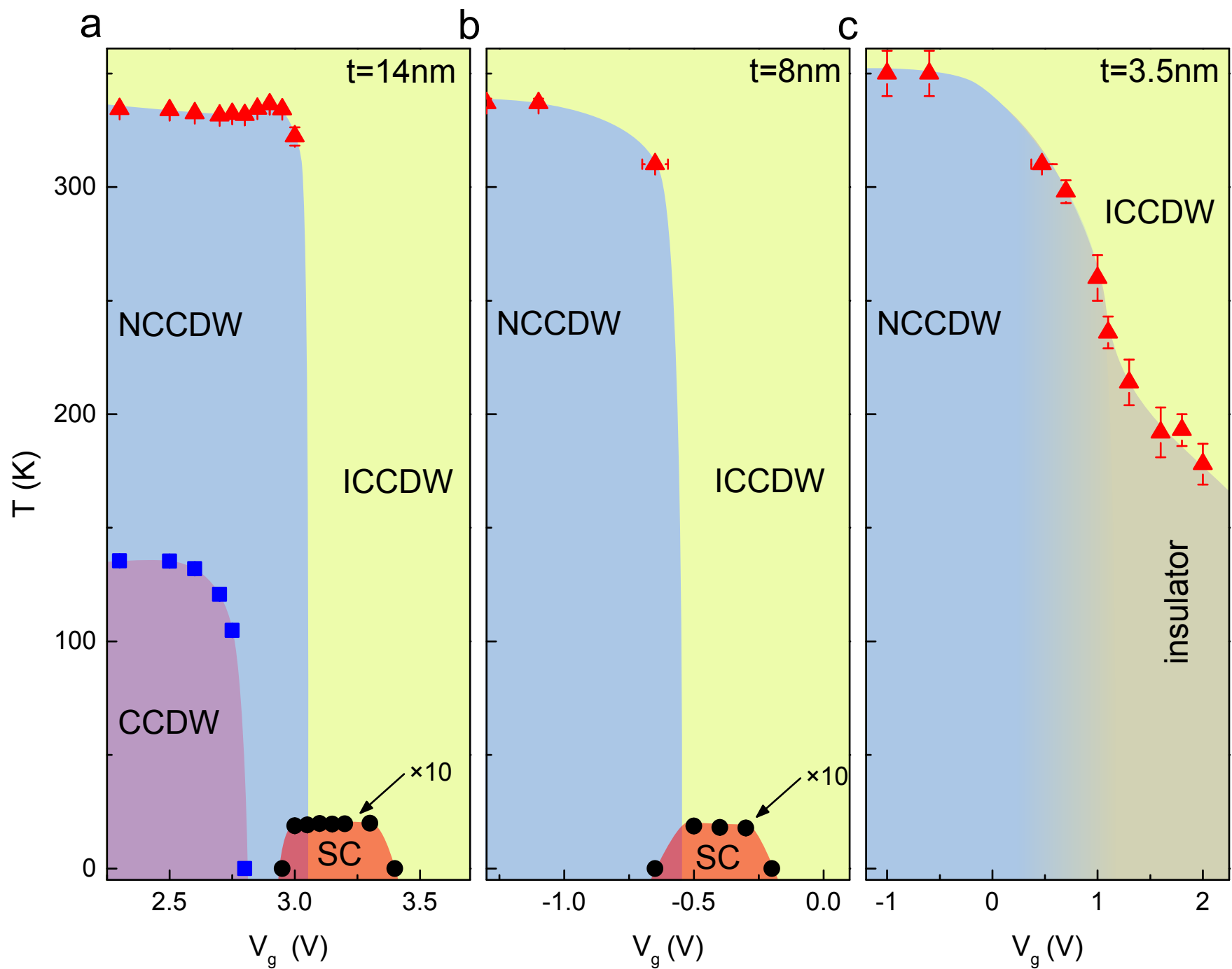

Fig. 5, Y. Yu et al

# Supplementary Information for
# Gate-tunable Phase Transitions in 1T-TaS$_2$

Yijun Yu, Fangyuan Yang, Xiu Fang Lu, Ya Jun Yan, Y. H. Cho, Liguo Ma, Xiaohai Niu, Sejoong Kim, Young-Woo Son, Donglai Feng, Shiyan Li, Sang-Wook Cheong, Xian Hui Chen and Yuanbo Zhang*

*Email: zhyb@fudan.edu.cn

## Contents

# I. Characterization of bulk 1T-TaS$_2$ single-crystal

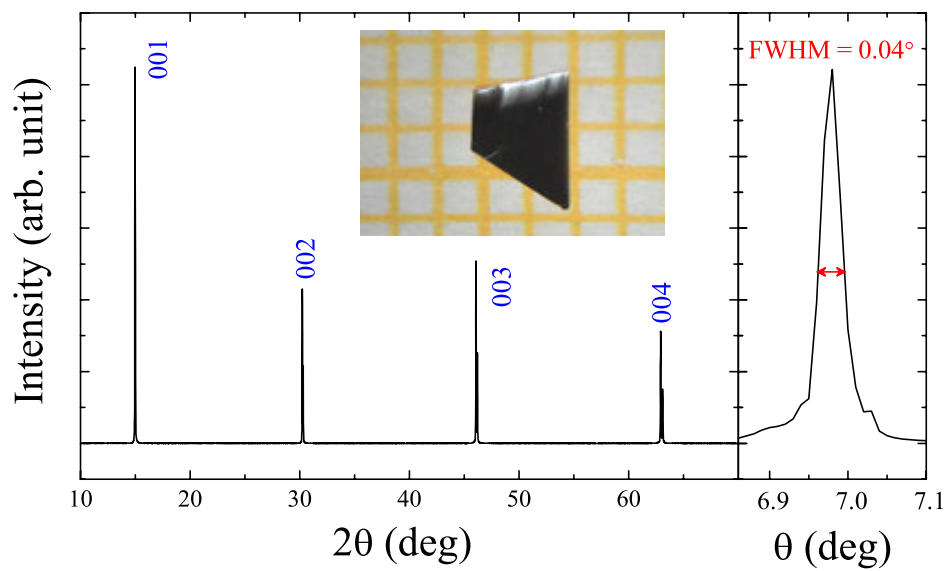

**Figure S1. XRD of bulk 1T-TaS$_2$.** Inset shows an optical image of a typical bulk crystal. The size of the squares in the background is 1mm×1mm. The 0.04° full width at half maximum (FWHM) of the diffraction peak indicates the high quality of the bulk crystal.

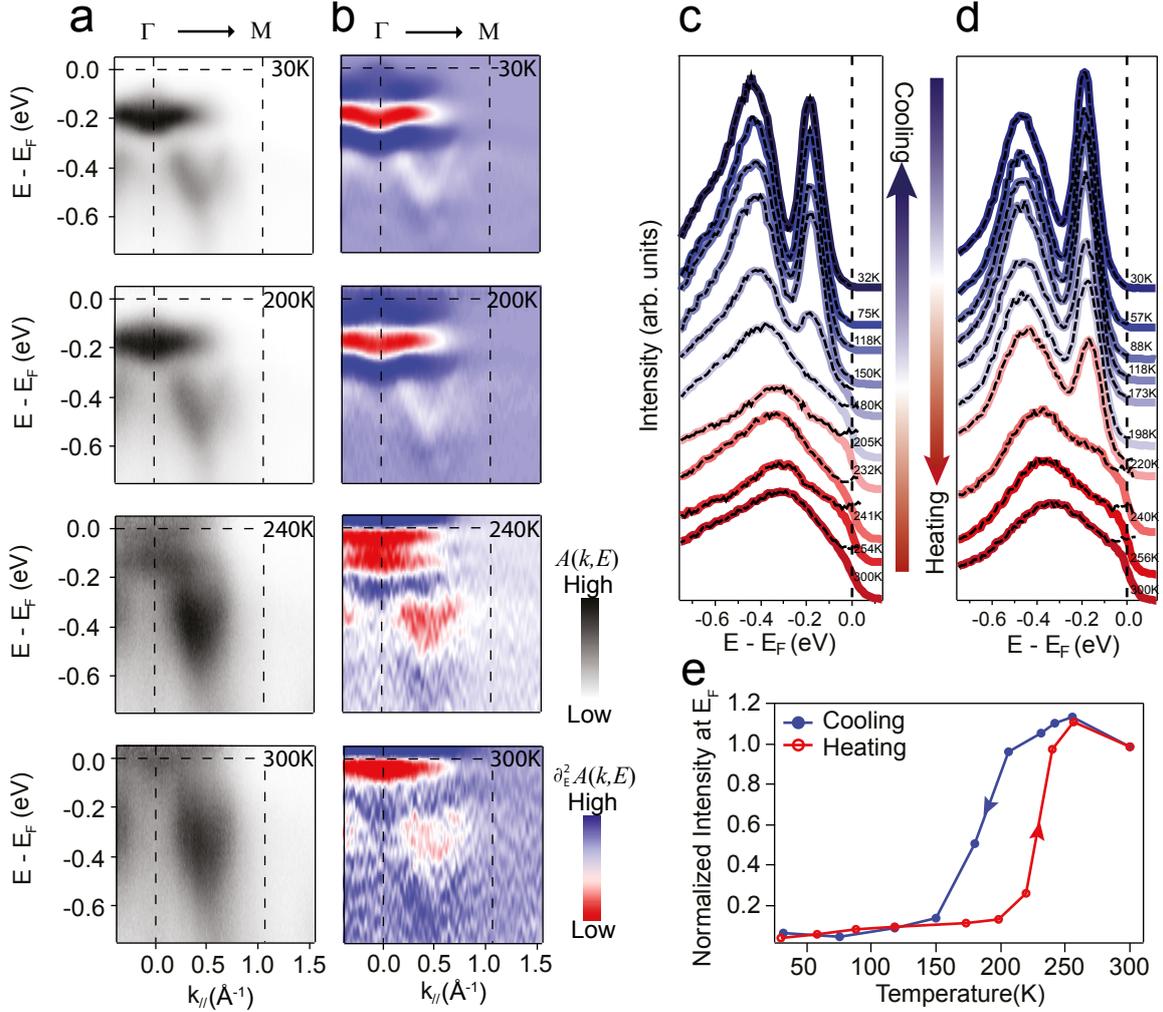

**Figure S2. ARPES measurements of bulk 1T-TaS$_2$. a,** Photoemission intensity along the Γ-M direction taken at increasing temperatures 30 K, 200 K, 240 K and 300 K. The intensities measured at 200 K, 240 K and 300 K are divided by Fermi-Dirac function at corresponding temperatures convoluted with a Gaussian with 7 meV FWHM. The horizontal and vertical dashed lines represent Fermi level and high symmetry points (Γ and M), respectively. **b,** The second derivatives of the photoemission intensity shown in **a** with respect to energy for better visualization of the band dispersion, which is consistent with previous study[10]. At low temperatures (30 K and 200 K), a Mott-Hubbard band at ∼ -200 meV is clearly resolved in the CCDW state. A flat band emerges at the Fermi level in the NCCDW state upon increasing the temperature (*T* = 240 K), and meanwhile the Mott-Hubbard band eventually disappears at high temperature (*T* = 300 K). **c** and **d,** Spectra taken at $k_{//}$ = 0.4 Å$^{-1}$ in Γ-M direction at varying temperatures where adjacent bands are well separated during cool-down and warm-up, respectively. Hysteresis was observed in the temperature-dependent spectra as a result of the first-order nature of the CCDW/NCCDW phase transition. The dashed lines are the normalized spectra after removing the temperature smearing due to Fermi-Dirac distribution. Intensity of the normalized spectra at the Fermi level vanishes at around 150 K during cool-down and emerges at 220 K during warm-up. **e,** The temperature dependence of the spectral weight of intensity photoemission in **c** and **d** measured at Fermi level. All data points are normalized to the value at 300 K. Hysteresis is observed in agreement with the transport data discussed in the main text. Meanwhile the intensity peak at ∼ -420 meV shifts toward Fermi energy during warm-up and moves away during cool-down, and similar hysteresis is also observed.

# II. Characterization of 1T-TaS$_2$ thin flakes

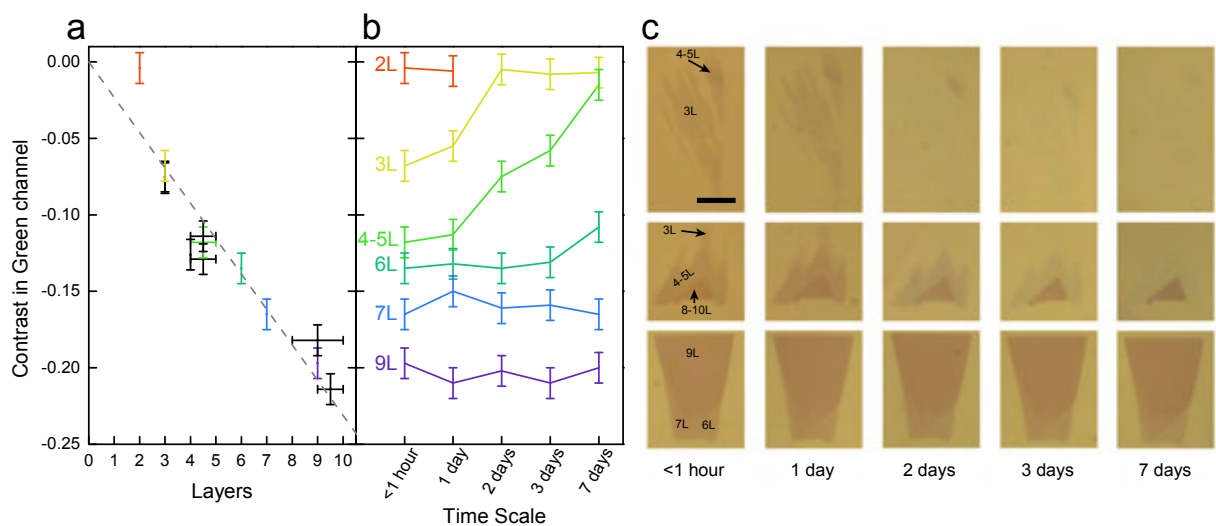

**Figure S3. Optical characterization of 1T-TaS$_2$ thin flakes and their time and thickness dependent degradation. a,** Optical contrast as a function of number of layers for freshly cleaved thin flakes. Contrast was extracted in the green channel of the image of the thin flakes[11]. The thickness, and correspondingly the number of atomic layers, was determined by AFM measurements. The optical contrast obtained this way correlates well with the thickness, and a good linear relation is found for samples thicker than 1.2 nm (2 layers). Such a well-behaved linear dependence is useful for an estimation of the sample thickness even without time-consuming AFM measurement. **b,** Optical contrast as a function of time period the sample exposed to air. The same flakes as in **a** are used in this study, and the curves are color-coded to match the data in **a**. **c,** The optical image of the flakes measured in **a** and **b**. Scale bar: 5 μm. We note that no degradation was observed for flakes thicker than 7 layers, and the contrast was stable even after 1 week in air. But discoloration was observed over time for samples below 6 layers, and the contrast decreases more quickly for thinner samples. We also note that the bilayer (see Fig. S5 for detail) sample shows a vanishing contrast at the very beginning, likely because it has already degraded before we could take a reliable optical image. This also explains why the contrast of the bilayer sample deviates from the linear behaviour shown in **a**.

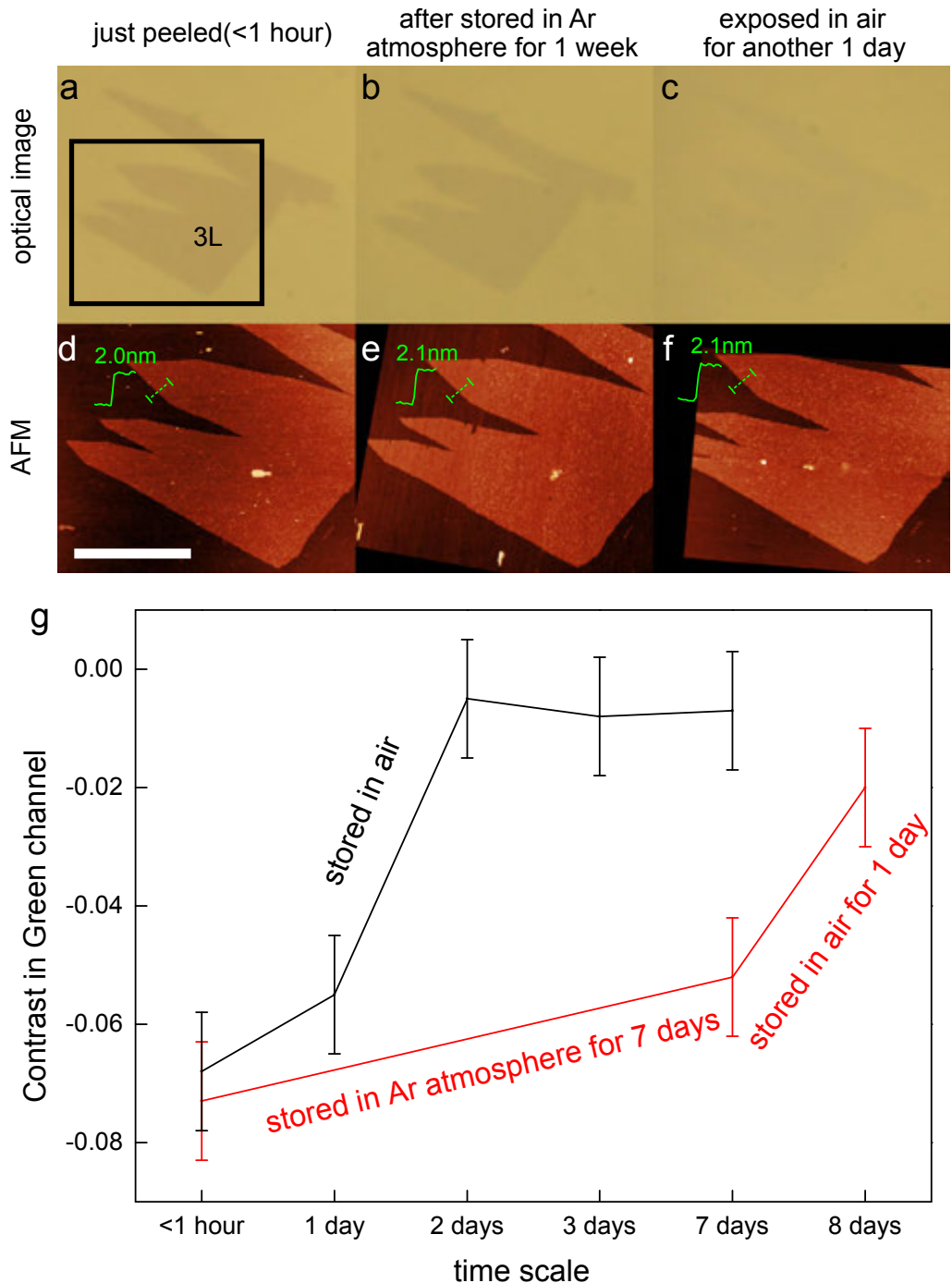

**Figure S4. Degradation of 1T-TaS$_2$ tri-layer in air and Ar atmosphere. a – c,** Optical images of the tri-layer thin flake right after cleaving, after 1 week in Ar, and after another one-day exposure to air, respectively. **d – f,** AFM images of the flakes shown in **a – c**. The AFM images are taken in the area marked by the square in **a**. The thickness of the sample is measured in the line profile shown on each image. No obvious change in thickness is detected during the sample degradation. Scale bar: 5 μm. **g,** Optical contrast observed over time in tri-layer samples stored in air and Ar. The red curve shows the contrast of a tri-layer flake stored in Ar and air over time (data extracted from **a – c**), and the black curve shows the contrast of another tri-layer sample stored in air for comparison. Keeping the thin-flake samples in Ar atmosphere significantly slows down the degradation. This observation supports our speculation that chemical reaction with H$_2$O or O$_2$ in air causes sample degradation, and is also consistent with previous finding that metallic transition metal dichalcogenides (TMD) are stable in Ar atmosphere[12].

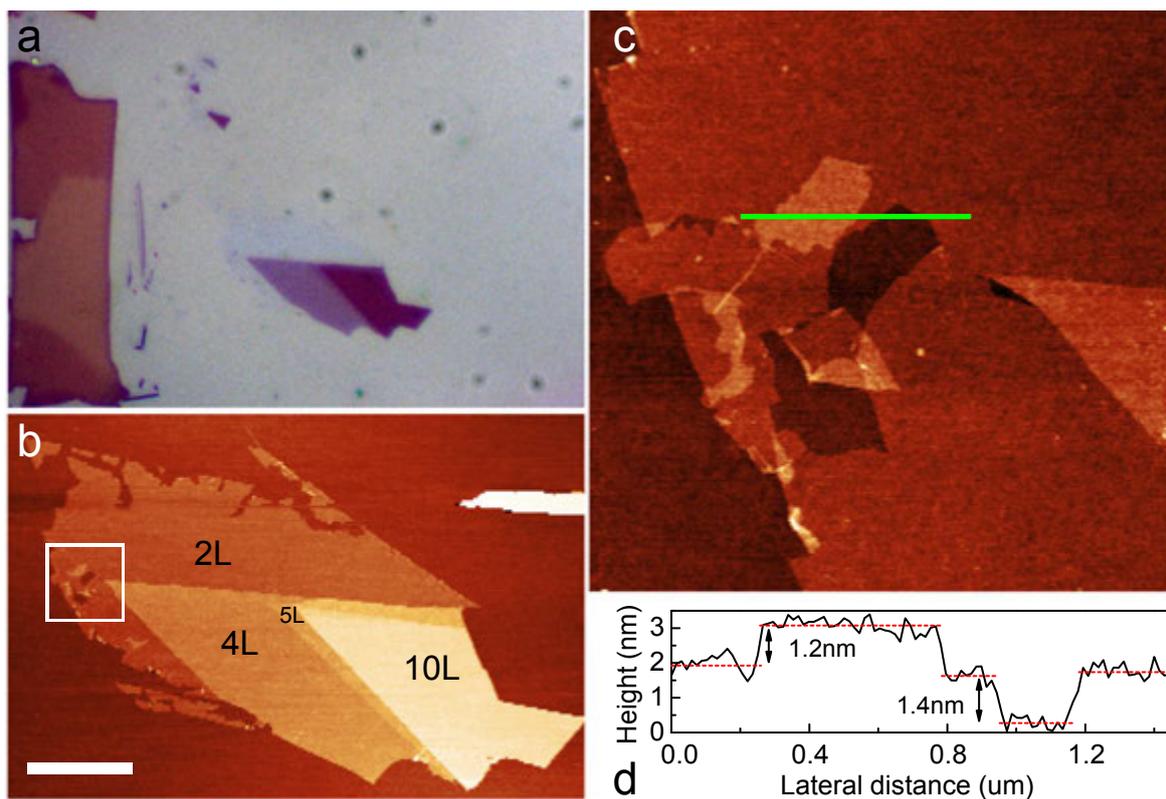

**Figure S5. Thickness characterization of a bi-layer flake. a,** Optical image of a bilayer flake with enhanced contrast. The vanishing contrast indicates that the flake is already degraded. **b and c,** AFM images of the same bi-layer flake shown in **a. c,** A zoomed-in view of the square marked in **b**. The scale bar in **b** is 5 μm. **d,** Line profile along the green line shown in **c.** An accurate measurement of the sample thickness (1.2 nm) is made possible by the sample's folding back on itself. It should be noted that monolayer flakes may have already been produced during exfoliation, but their vanishing contrast due to sample degradation may make them difficult to find.
6

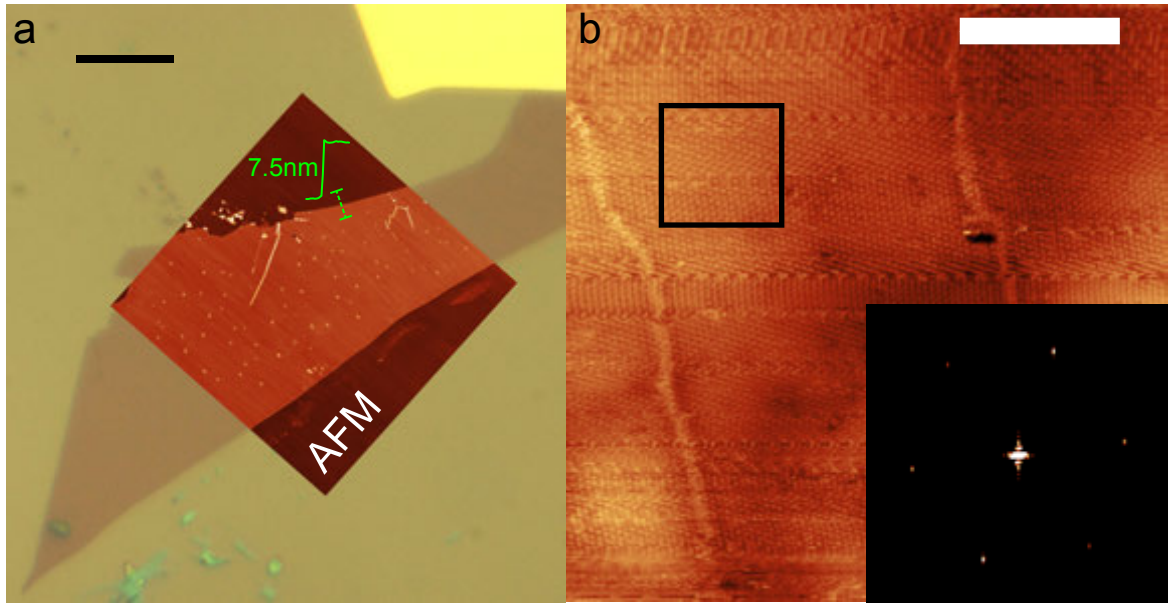

**Figure S6. STM characterization of a 7.5-nm-thick flake. a,** Optical image of a 7.5-nm-thick flake. Overlapped on top of the graph is an AFM topography image of the same sample. Sample thickness is determined from the line profile (7.5 nm). Scale bar: 10 μm. **b,** STM image of the same sample shown in **a**. Scale bar: 20 nm. STM measurements were carried out on the sample without any pre-treatment in vacuum, and still we found large clean areas on the surface showing CDW superstructure and domain boundaries. The inset shows the Fast Fourier Transformation (FFT) of the area marked by the black square. FFT patterns with six-fold symmetry due to the CDW ordering are clearly resolved. Our experiment indicates that air exposure does not damage the sample surface for this 7.5-nm-thick sample.

# III. Gate-induced NCCDW/ICCDW transition in 1T-TaS$_2$ iFET

We attribute the gate-induced resistance modulation observed in Fig. 1d to the NCCDW/ICCDW phase transition in 1T-TaS$_2$. A direct evidence comes from the measurement of $R$ as a function of $V_g$ at fixed temperatures between 300 K and 360 K (Fig. S7a). To avoid complications from hysteresis, here only the data from the up-sweep of $V_g$ is displayed, and the temperature is varied in one direction (from 300 K to 360 K). The gate-induced resistance drop is clearly visible at $V_g \sim 2$ V when temperature is held at 300 K. As the temperature is increased, the resistance drop is pushed to lower $V_g$. At 350 K, where the NCCDW/ICCDW phase transition is about to emerge in a pristine sample, the resistance drop in the gated sample is found at $V_g$ close to 0 (~ 0.4 V, Fig S7a, green curve). This observation indicates that the gate-induced resistance drop is a result of the same NCCDW/ICCDW transition induced by temperature variation. Indeed, if we extract the value of sample resistance at $V_g = 0$ V at each temperature, the data (Fig. S7b, red squares) follow closely the behaviour of a pristine sample at the NCCDW/ICCDW phase transition (Fig. S7a, black line). The phase diagram of 1T-TaS$_2$ as function of temperature and $V_g$ is summarized in Fig. S7c. It is clear from the phase diagram that both gate doping and temperature variation can effectively modulate the same NCCDW/ICCDW phase transition.

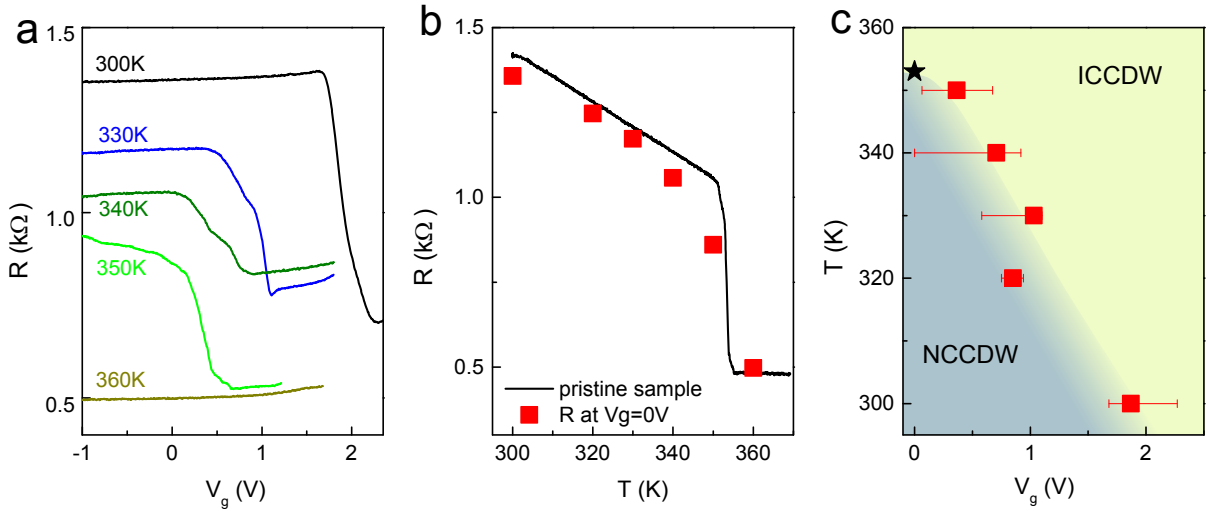

**Figure S7. Gate-induced phase transition in 1T-TaS$_2$ iFET. a,** Resistance of a 5-nm-thick sample as a function of $V_g$ measured at varying temperatures. Only the up-sweep of $V_g$ is shown and temperature is varied in one direction (from 300 K to 360 K). The sudden resistance drops signify the NCCDW to ICCDW phase transition. **b,** The temperature-dependent resistance of pristine 1T-TaS$_2$ (black line) and the same sample subjected to ionic gating in an iFET (red squares). The data on the gated sample are extracted from temperature-dependent $V_g$ sweeps (shown in **a**) at $V_g = 0$ V. **c,** The temperature - $V_g$ phase diagram of the same sample. The star indicates the transition temperature of the same sample before the application of solid electrolyte.

# IV. Gate-controlled intercalation in 1T-TaS$_2$ iFET

## 1. Extreme charge doping induced by gate-controlled intercalation

To further illustrate the extreme doping level of the gate-controlled intercalation, we plot the carrier density as a function of $V_g$ (Fig. 4d in main text) in linear scale (Fig. S8b). We see that the absolute carrier density induced by the gate can be 2 orders of magnitude higher than the doping level expected from electron double layer (EDL) surface gating (Fig. S8b, red dashed line). More importantly, the relative change in the carrier density induced by the gate is also much larger than that expected from EDL gating (Fig. S8b). We emphasize that such a large modulation (carrier density changed by factor of 4 between $V_g$ = 3.15 V and $V_g$ = 3.6 V) is induced when the sample is in a metallic state with high 2D carrier concentration (> $10^{16}$ cm$^{-2}$). The EDL surface gating cannot possibly provide such high doping levels to induce the effects observed here (in special cases such as VO$_2$, the mechanism of charge doping is still debated[1,2]).

## 2. Homogeneity of intercalated samples

As our samples are typically micron-size and buried underneath the electrolyte, it is very difficult to inspect their crystal structure and homogeneity using traditional diffraction and scanning probe techniques. However, valuable information about the homogeneity of gated samples can still be obtained from our transport data. First of all, we point out that all phase transitions (including the CDW transitions as well as the superconducting transition, such as those shown in Fig. 4a) remain sharp, i.e. without broadening or multiple steps, under gate modulation. This observation indicates homogeneous doping across the entire sample.

Secondly, we are able to infer good sample homogeneity from the behaviour of sample resistance in the normal state. Fig. S8c displays the zoomed-in view of the low temperature part of the sample resistance shown in Fig. 4a. Above $V_g$ = 3.4 V, where the sample turns into a normal metal, its resistance curves show no detectable change at superconducting transition temperature. The complete disappearance of any sign of the superconductivity indicates a homogeneous sample, because if only part of the sample remains superconductive, we would have observed a drop in its resistance at the transition temperature.

## 3. Hysteresis of 1T-TaS$_2$ iFET during gate sweeps

As is typical for ion-based gate modulation[1], large hysteresis is present in our 1T-

TaS$_2$ iFETs during gate voltage sweeps. Fig. S8d and Fig. S8e show two typical hysteresis loops with LiClO$_4$/PEO solid electrolyte from two different batches. Although the first-order NCCDW/ICCDW phase transition is expected to cause hysteresis intrinsic to the sample, we found that external factors also contribute to the hysteretic gating behaviour of the iFET. First, the gating capability was found to be very sensitive to the LiClO$_4$ concentration in PEO. A slight difference due to experimental uncertainty causes significant change in the hysteresis behaviour. Second, the gating capability of the electrolyte gradually decreases over time, and loses its gating ability after several weeks, possibly due to lithium ions losing their mobility over time. To avoid the aging effect, we only used electrolyte freshly made within a week. Third, temperature also plays an important role in the gating behaviour of the solid electrolyte due to the temperature-dependent mobility of the lithium ions in the electrolyte.

Due to factors listed above, the absolute value of $V_g$ is not a good indicator for the extent of intercalation or charge doping, even though the same qualitative gating behavior is observed in more than 41 samples that we have measured. Instead we used the charge carrier density determined by Hall measurement as an independent measure of the charge doping level (Fig. 4d).

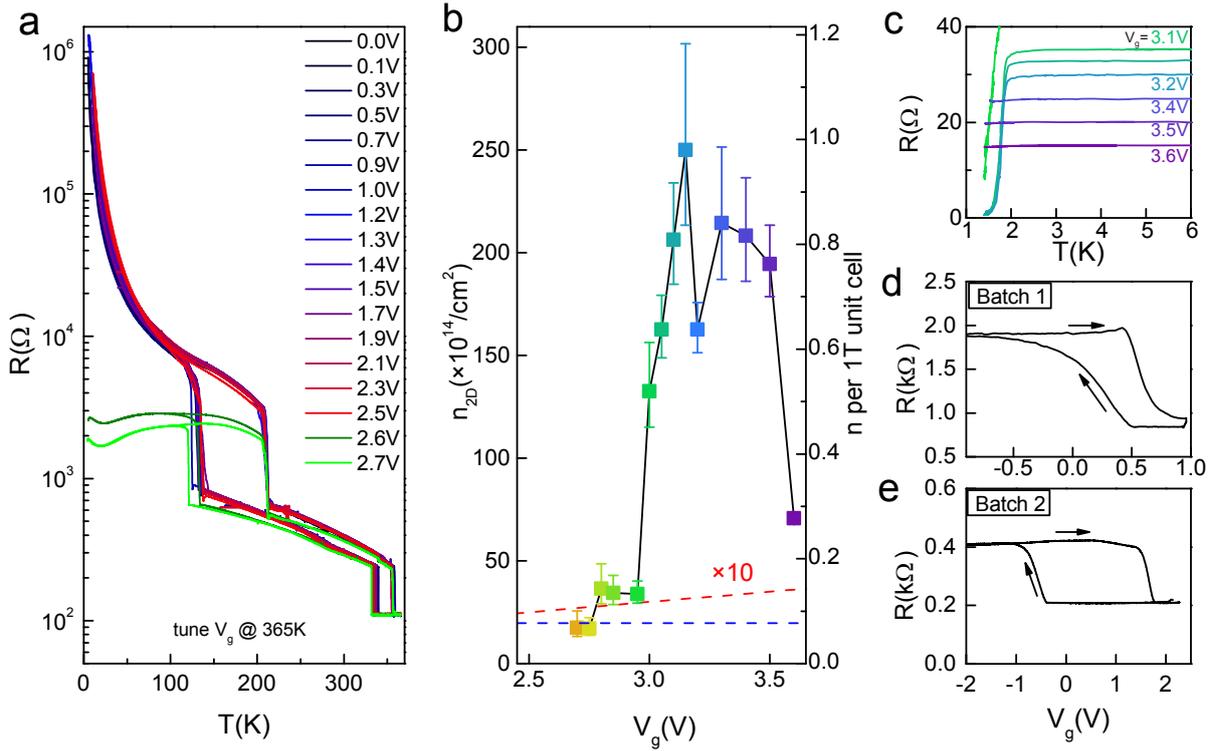

**Figure S8. The gating behavior of iFET. a,** Resistance as a function of temperature for the 14-nm-thick sample shown in Fig. 4a under varying $V_g$. **b,** Carrier density as a function of $V_g$ in Fig. 4d, shown here in a linear scale. The red dashed line indicates the upper limit of EDL doping capability (~$1\times10^{14}$cm$^{-2}$V$^{-1}$, see ref. 13). The blue dashed line indicates the doping level corresponding to 1 *e* per David-star. **c,** The zoomed-in view of the low temperature part of Fig. 4a. **d** and **e**, Examples of observed hysteresis in sample resistance during gate sweeps. The two samples shown here used electrolytes from two different batches with nominally the same lithium ion concentration.

## V. 1T-TaS$_2$ iFET with other ion species

More evidences supporting the gate-controlled intercalation mechanism are gained by applying the principle of iFET to other ion species. Specifically, we have used i) Bis(trifluoromethane)sulfonimide lithium salt dissolved in PEO (Li-TFSI/PEO), ii) KClO$_4$/PEO, iii) N,N-diethyl-Nmethyl-N-(2-methoxyethyl)ammonium bis(trifluoromethylsulphonyl)imide (DEME-TFSI) gated at 225 K, and iv) DEME-TFSI gated at 290 K, along with v) LiClO$_4$/PEO and vi) PEO used as control experiments. The results are shown in Fig. S9. There are three main points to notice. First, substituting ClO$_4^{1-}$ in the electrolyte with other anion species does not appreciably affect the gate doping, as demonstrated by the blue and red curves (corresponding to i and v, respectively) in Fig. S9a. Second, replacing the lithium ions in the electrolyte with other, larger cation species dramatically alters the iFET behaviour (dark green curve compared with red curve in Fig. S9a, corresponding to ii and v, respectively). We attribute the change in the iFET behaviour to possible structural damage caused by the large potassium ions between the atomic layers of 1T-TaS$_2$, which is also a common problem found in alkali-ion batteries[3].

The third intriguing evidence comes from a direct comparison between gate-controlled intercalation and EDL gating on a 7-nm-thick 1T-TaS$_2$ flake. We used ionic liquid DEME-TFSI as the electrolyte, and swept $V_g$ at 225 K – exactly the same condition where EDL has been found to forms at the sample surface[4,5]. Only 3% change in the sample resistance is observed for $V_g$ up to 6 V ($V_g$ sweep below 2.3 V is shown in Fig. S9b), with no signs of CDW phase transitions. The null effect is not surprising given the fact that 1T-TaS$_2$ crystal in the NCCDW state is a metal with high carrier concentration, which can not be easily tuned by surface gating. Increasing the temperature to 290 K, however, dramatically changes the gating behaviour of the DEME-TFSI ion liquid: the sample resistance rises sharply by over one order of magnitude at $V_g \sim$ 1.5 V, and slowly comes down beyond that (black curve, Fig. S9b). The observed behaviour is a drastic departure from EDL gating, but shows remarkable similarity to hydrazine intercalation in bulk 1T-TaS$_2$ over time (Fig. S9b inset, reproduced from ref. 6), as well as potassium ion intercalation shown in Fig. 9a. Our observation indicates that, much like hydrazine and potassium, DEME$^+$ is also able to be intercalated into 1T-TaS$_2$, causing structural damage. The similarity among the three anion spices may be explained by the fact that they are all large enough to decouple 1T-TaS$_2$ layers and they all act as electron donors.

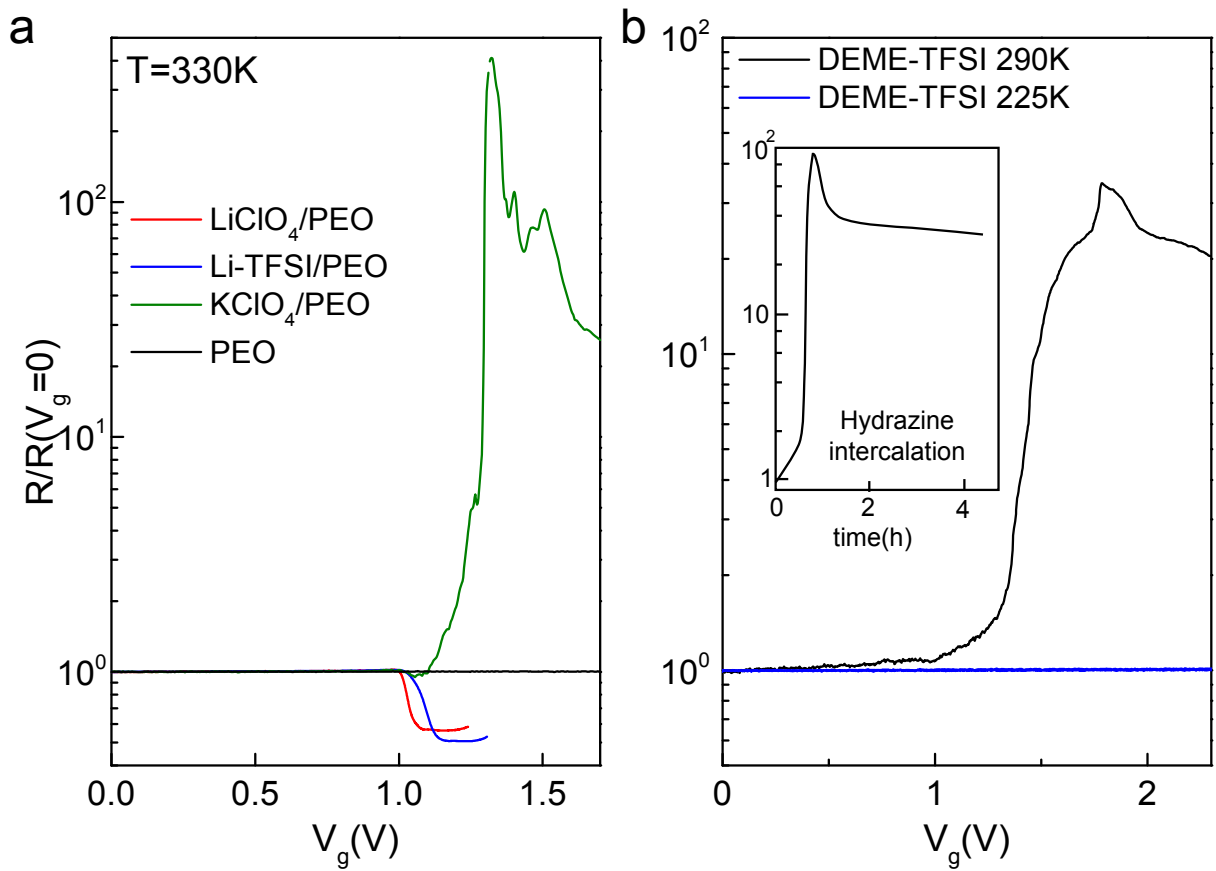

**Figure S9. Comparison of iFET gating behaviour with various electrolytes. a,** The gating behaviour of 1T-TaS$_2$ iFET at 330 K using LiClO$_4$/PEO (red), Li-TFSI/PEO (blue), KClO$_4$/PEO (green), and PEO (black) as gate medium. **b,** The gating behaviour of 1T-TaS$_2$ iFET using ionic liquid DEME-TFSI. Data are taken from the same sample at 290 K (black) and 225 K (blue). Inset: Resistance of bulk 1T-TaS$_2$ subjected to hydrazine intercalation recorded as a function of intercalation time (reproduced from ref. 6).

## VI. Temperature hysteresis of CDW phase transitions

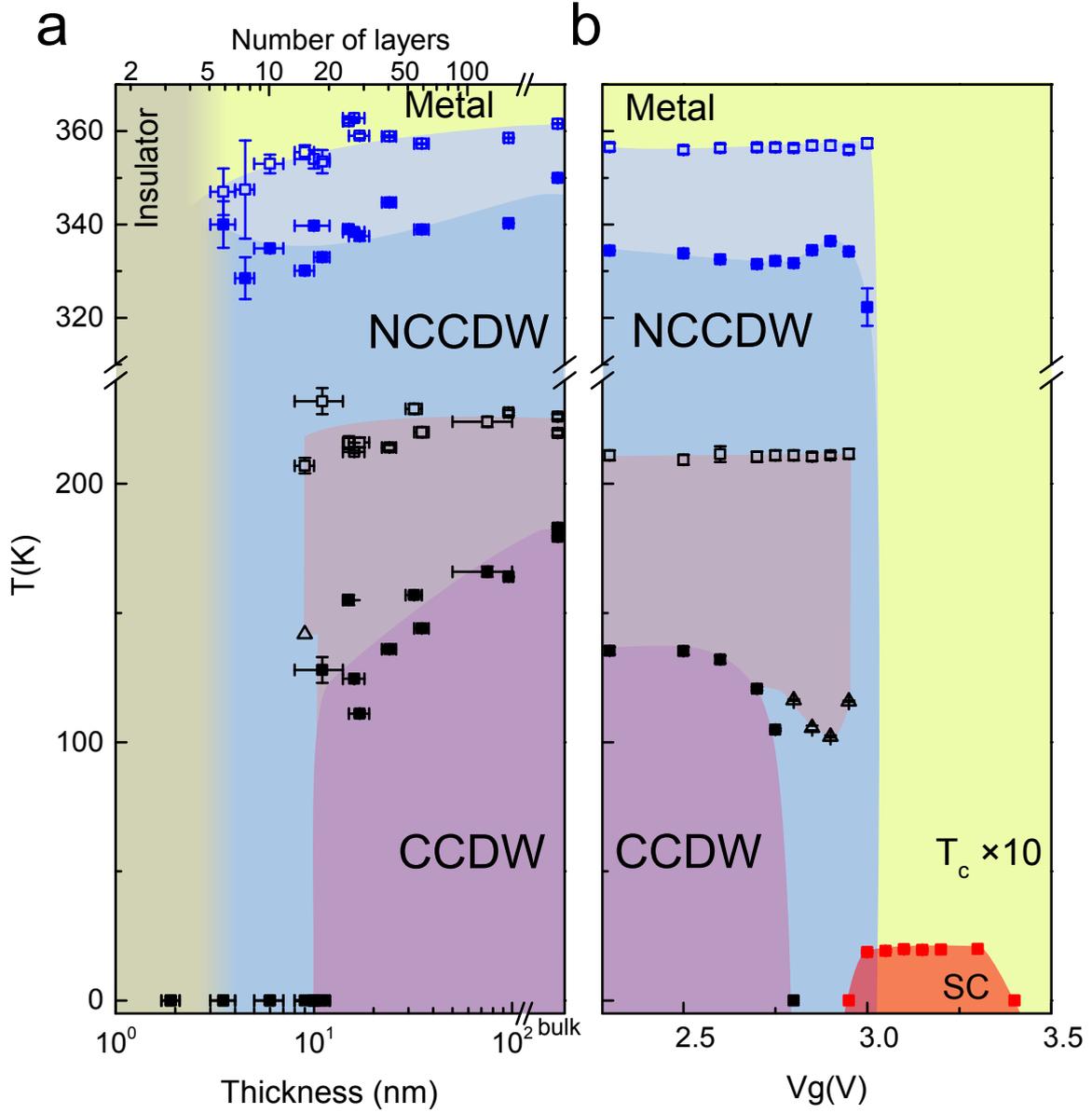

**Figure S10. Hysteresis of the CCDW and NCCDW transitions.** Hysteresis is observed at the CDW phase transitions during temperature sweeps due to the fact that the transitions are first-order. The transitions for both up-sweep and down-sweep were recorded in **a,** thickness-temperature and **b,** gate-temperature phase diagrams here (The same transitions are also shown in Fig. 3b and 5a, but only the down-sweep transitions are displayed for simplicity). We note that there exists a region (8 nm < thickness < 12 nm in **a,** and 2.6 V < Vg < 2.9 V in **b**) where CCDW phase is absent when cooling down, but shows up during warm-up. This phenomenon was also observed in 1T-TaS$_2$ samples under high pressure[14], and its origin is still not clear at this moment. Recent experiment on bulk 1T-TaS$_2$ excited by femtosecond laser showed transport behaviour similar to that seen here[15], which may provide clues for understanding the observed phase.

## VII. Upper critical field of the superconducting state in doped 1T-TaS$_2$

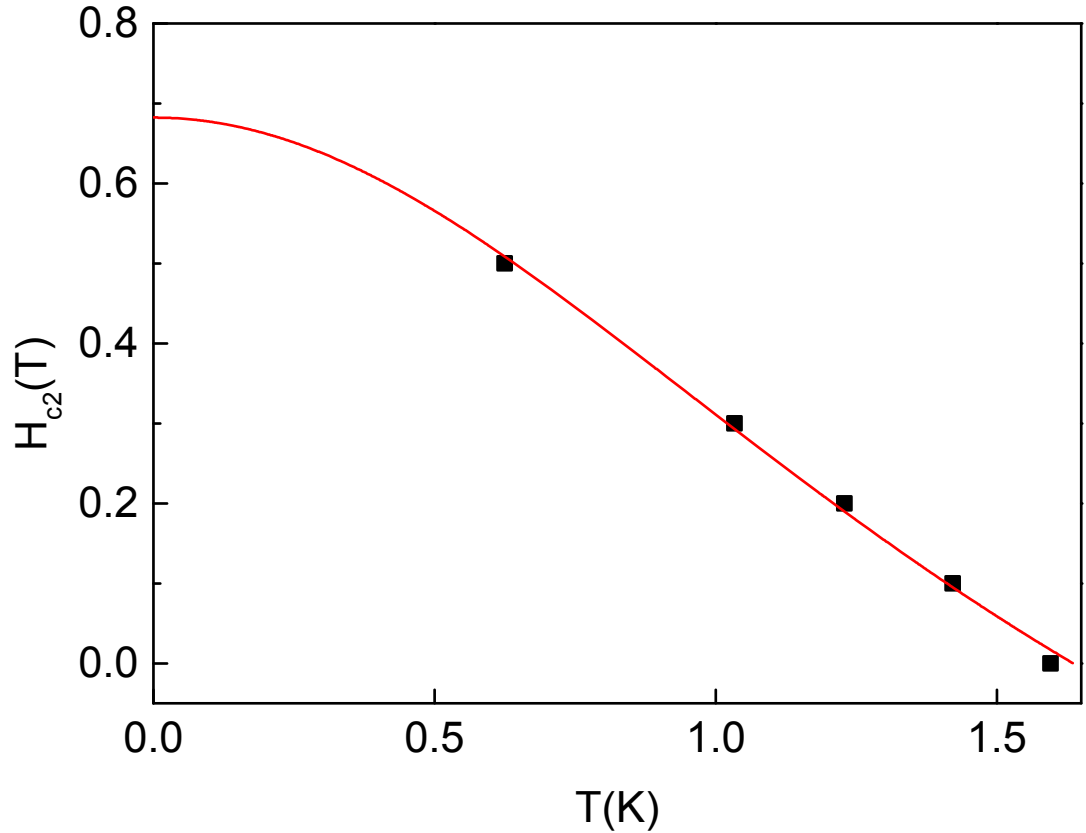

**Figure S11. Upper critical field of doped 1T-TaS$_2$ in the superconducting state.** The measured values of $T_c$ and $H_{c2}$ are shown as black squares. Here $T_c$ is defined as the temperature at 10% drop from the normal state resistance. The data are fitted with Ginzburg-Landau formula for type-II superconductors (red curve). The good fit indicates that the doped 1T-TaS$_2$ is a type-II superconductor. An upper critical field of $H_{c2} \sim 0.7$ T at zero temperature is obtained from the fitting.

# VIII. Calculations on Li intercalation of 1T-TaS$_2$

To investigate doping effects from intercalated Li atoms, we performed density function theory (DFT) calculations by adopting the PBE generalized gradient approximation[7] for exchange-correlation functional and the norm-conserving pseudo-potentials[8]. We used the plane-wave DFT code QUANTUM-ESPRESSO[9] with a cutoff of 55 Ry and a smearing temperature of 0.01 Ry. A calculated lattice constant for a unit cell of pristine 1T-TaS$_2$ is 3.28 Å.

We performed a relaxation calculation by placing one Li atom in a 4×4×1 super-cell of 1T-TaS$_2$ to find the equilibrium position of Li atoms intercalated in 1T-TaS$_2$. The minimum of the total energy was generally reached when the intercalated Li atom is located on the line connecting Ta atoms that lie on the same sites of adjacent layers. We next calculated electron doping densities of Li-intercalated 1T-TaS$_2$ structures for five cases: 1) one Li atom in the 4×4×1 super-cell, 2) two Li atoms in the 4×4×1 super-cell, 3) one Li atom in the 4×4×2 super-cell, 4) two Li atoms located in the same interlayer space of the 4×4×2 super-cell, and 5) two Li atoms located in neighboring interlayer spaces. Throughout these calculations the interlayer distance was fixed at $c = 5.72$ Å because the additional forces on the layers with Li intercalations are negligible. Fig. S12a and S12b correspond to case 1) and Fig. S12c-S12f correspond to cases 2)-5) described above.

For each of the configurations shown in Fig. S12, we calculated the charge density of 1T-TaS2 with and without Li intercalation (denoted as $\rho_{TaS2-Li}$ and $\rho_{TaS2}$, respectively), and the charge density of Li without 1T-TaS$_2$, $\rho_{Li}$. By integrating along $x$ and $y$ in-plane directions, the charge density along the $z$ axis is obtained:

$$\Delta\rho(z) = \int dxdy \left[\rho_{TaS_2-Li}(x,y,z) - \rho_{TaS_2}(x,y,z) - \rho_{Li}(x,y,z)\right].$$

Using $\Delta\rho(z)$, we calculated the net charge transfer to the 1T-TaS$_2$ region, which is defined as $\Delta Q = \int_{-z_0}^{z_0} dz \Delta\rho(z)$. Here we set $\pm z_0$ to be points where $\Delta\rho(z) = 0$ so $\Delta\rho(z)$ changes its sign around these points. Charge-doping densities per layer for the five cases are summarized in Fig. S12. We note that for all five cases the Li intercalation does not change the energy band-structures, but instead pushes up the Fermi energy by an amount determined by the doping level. From Fig. S12g, we found that the amount of charge transfer between Li atom and 1T-TaS$_2$ lattice is proportional to the number of intercalated Li atoms, and each Li atom donates ~ 0.2 $e$ to the host 1T-TaS$_2$ on average. Such a high doping level is achieved without causing structural instabilities. We note that the charge transfer is calculated for an ideal, undistorted 1T-TaS$_2$ lattice, and the amount of transfer may be even higher if the periodic lattice distortion associated with CDW states is taken into account.

We further estimated the maximum intercalation density obtainable before the structure becomes unstable. For the case where one Li atom is intercalated into one 1×1 unit cell of 1T-TaS$_2$, we relaxed the inter-layer distance, and found that the minimum of the total energy was reached when c increases from 5.72 Å (corresponding to zero intercalation) to 6.34 Å. A maximum electron doping of 4.0×10$^{14}$ $e$/cm$^2$ was achieved in this case. We then relaxed the 1×1 unit cell of 1T-TaS$_2$ intercalated by two Li atoms. In this case a finite interlayer distance that minimizes the total energy cannot be obtained - as the interlayer distance increases, the total energy keeps decreasing. Our results imply that the 1T-TaS$_2$ structure intercalated by two Li atoms per unit-cell is energetically unstable, and suggest that the maximum intercalation is one Li atom per unit cell.

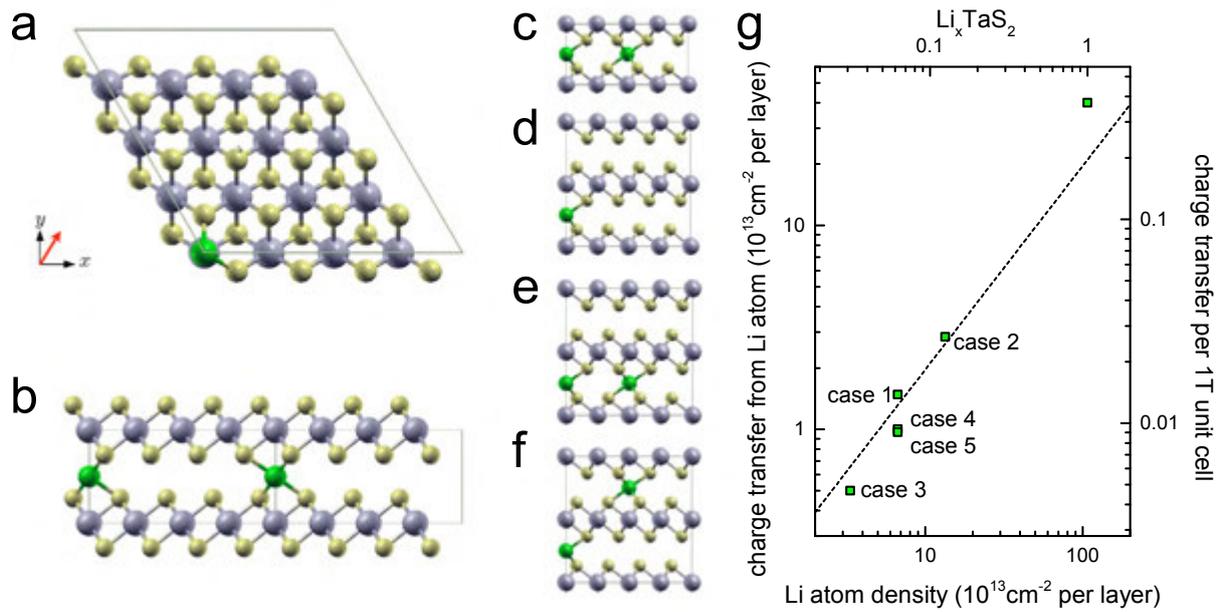

**Figure S12. Atomic models for Li intercalated 1T-TaS$_2$. a,** Top view of a 4×4×1 super-cell of 1T-TaS$_2$ atomic model with one intercalated Li atom. The black parallelogram denotes the 4×4 super-cell in the x-y plane. The blue, yellow and green spheres represent Ta, S and Li atoms, respectively. **b,** Side view of a along the red arrow shown in a. One Li atom is intercalated per 4×4×1 super-cell. **c,** Side view of a 4×4×1 super-cell. Two Li atoms are intercalated per super-cell. **d,** Side view of a 4×4×2 super-cell. One Li atom is intercalated per super-cell. **e** and **f**, Stable atomic configurations for two intercalated Li atoms per 4×4×2 super-cell in the same inter-layer and in different inter-layer space, respectively. g, Calculated charge transfer as a function of Li atom density in intercalated 1T-TaS$_2$. From the slope of the line fit (broken line), we estimate that each intercalated Li atom, on average, donates ~ 0.2 electrons to the host 1T-TaS$_2$.